%% file: main.tex
\documentclass[conference]{IEEEtran}
\IEEEoverridecommandlockouts
\usepackage{cite}
\usepackage{amsmath,amssymb,amsfonts}
\usepackage{algorithmic}
\usepackage{graphicx}
\usepackage{textcomp}
\usepackage{xcolor}
\usepackage{color}
\usepackage{url}
\usepackage{soul}

\newboolean{showcomments}
\setboolean{showcomments}{true}
\ifthenelse{\boolean{showcomments}}
{

\newcommand{\mynote}[2]{
    {\color{#1}Note: #2}
}
\newcommand{\note}[2]{
    {\color{#1}Yulong: #2}
}
\newcommand{\Dirk}[2]{
    {\color{#1}\fbox{\begin{minipage}{0.8\linewidth}Dirk: #2\end{minipage}}}
}
\newcommand{\Jeff}[2]{
    {\color{#1}\fbox{\begin{minipage}{0.8\linewidth}Jeff: #2\end{minipage}}}
}
} 
{
\newcommand{\mynote}[2]{}
\newcommand{\note}[2]{}
\newcommand{\Dirk}[2]{}
\newcommand{\Jeff}[2]{}
} 

\newcommand{\dirk}[1]{\Dirk{blue}{#1}}

\def\BibTeX{{\rm B\kern-.05em{\sc i\kern-.025em b}\kern-.08em
    T\kern-.1667em\lower.7ex\hbox{E}\kern-.125emX}}

\begin{document}

\title{Secure Web Objects: Building Blocks for \\Metaverse Interoperability and Decentralization\\
}

 \author{\IEEEauthorblockN{Tianyuan Yu}
 \IEEEauthorblockA{\textit{UCLA} \\
USA \\
tianyuan@cs.ucla.edu}
\and
\IEEEauthorblockN{Xinyu Ma}
\IEEEauthorblockA{\textit{UCLA} \\
USA \\
xinyu.ma@cs.ucla.edu}
\and
\IEEEauthorblockN{Varun Patil}
\IEEEauthorblockA{\textit{UCLA} \\
USA \\
varunpatil@cs.ucla.edu}
\and
\IEEEauthorblockN{Yekta Kocaogullar}
\IEEEauthorblockA{\textit{UCLA} \\
USA \\
ykocaogullar@g.ucla.edu}
\and
\IEEEauthorblockN{Yulong Zhang}
\IEEEauthorblockA{
\textit{HKUST (Guangzhou)}\\
China \\
yzhang893@connect.hkust-gz.edu.cn}
\and
\IEEEauthorblockN{Jeff Burke}
\IEEEauthorblockA{\textit{UCLA REMAP} \\
USA \\
jburke@remap.ucla.edu}
\and
\IEEEauthorblockN{Dirk Kutscher}
\IEEEauthorblockA{
\textit{HKUST (Guangzhou)}\\
China \\
dku@hkust-gz.edu.cn}
\and
\IEEEauthorblockN{Lixia Zhang}
\IEEEauthorblockA{\textit{UCLA} \\
USA \\
lixia@cs.ucla.edu}
}

\maketitle

\begin{abstract}
This position paper explores how to support the Web's evolution through an underlying data-centric approach that better matches the data-orientedness of modern and emerging applications. We revisit the original vision of the Web as a hypermedia system that supports document composability and application interoperability via name-based data access. We propose the use of \emph{secure web objects} (SWO), a data-oriented communication approach that can strengthen security, reduce complexity, centrality, and inefficiency, particularly for Metaverse and other collaborative, local-first applications.
SWO are application-defined, named, and signed objects that are secured independently of their containers or communications channels, an approach that leverages the results from over a decade-long data-centric networking research. This approach does not require intermediation by aggregators of identity, storage, and other middleware or middlebox services that are common today. We present a brief design overview, illustrated through prototypes for two editors of shared hypermedia documents: one for 3D and one for LaTeX. We also discuss our findings and suggest a roadmap for future research.
\end{abstract}


\input{1-intro}

\input{2-path}

\input{3-prototype}
\input{4-insights}
\input{5-agenda}

\input{relatedwork}

\input{conclusion}
\input{8-acks}


\bibliographystyle{IEEEtran}
\bibliography{reference}

\end{document}

%% file: 1-intro.tex
\section{Introduction}

The Web was conceived as a hypermedia information system enabling the publishing of Web objects and access to them by names (URIs). 
From a technical computer networking perspective, the `Metaverse' can be considered an advanced Web application.\footnote{Even game engine concepts for the Metaverse access the underlying content via standard web protocols in most cases.} 
In terms of user experience, it offers new spatialized immersive experience beyond the classic 2D browser interface. Technically, this is enabled by leveraging advances in 3D rendering, virtual and augmented reality, and by the combining existing Web technologies (protocols, media types) and interactive communication technologies, such as WebRTC
 \cite{kutscher2023statement,gonzalez2013virtual,nevelsteen2018virtual}.

On the other hand, Metaverse applications also offers us an opportunity to reexamine the Web and its underlying network foundations in a more principled way.
The current designs and deployment models typically treat the Metaverse as an overlay application with corresponding infrastructure dependencies.
This study follows a new direction introduced in \cite{kutscher2023statement} and  
envisions a fundamentally information-centric system in which applications engage in the exchange of granular 3D content objects, context-aware integration with the physical world, and other Metaverse-relevant services.

Today, many Metaverse applications  are intrinsically data-oriented. For example, multimedia document editors work with what can be viewed as catalogs of different types of media objects. Video streaming can be viewed more generally as a system of continuous production and consumption of media objects. Contemporary AI applications revolve around a variety of data types, such as training, models, and vector storage.  Emerging Metaverse systems work with interwoven collections of 3D scenes that organize media objects, which can be created, consumed, modified, and re-organized in new contexts.

What these systems have in common are concepts of  data objects accessed via namespaces, often expressed as URIs. 
However, the current Web stack effectively supports only a limited approximation of the original hypermedia concepts.
Data-centric behavior is supported at the application layer and within applications, but not at lower layers or as a means for interoperability between applications.  
There, a client-server communication approach is dominant; replacing single servers with centrally secured cloud resources has actually helped cement silos around data object collections. Dependence on channel-oriented security solutions, such as TLS, results in a lack of simple mechanisms to efficiently access shared data from peers or across domains, especially in local-first communication scenarios. 
Client-server connection-based communication patterns do not correspond well to the data-oriented nature of the applications they support, hence the popularity of a variety of messaging and middleware solutions that implement sit atop Web protocols and within cloud silos.  While data-centric in facing applications, these solutions typically rely on centralized and online services for authentication, rendezvous, and other key services. 

As a result, while such contemporary approaches have been successful in turning the Web into an application platform, they have moved away from the original vision of an interoperable hypermedia information system, even as the applications themselves become more data-centric. Secure data exchange \emph{across} existing Metaverse-style platforms is cumbersome, if available at all.  
As a simple thought experiment of interactive, object-level interoperability in Metaverse-style applications,
if one imagines that a teapot owned by one user in a Metaverse platform pours (virtual) tea into a cup owned by someone else on another platform, can each object and action be independently secured? Such cross-platform, cross-owner interaction is a simple building block of applications that is often invoked in shared virtual experiences.  This poses the challenge of supporting \textbf{secure interaction between objects} owned by different users and existing on different platforms, rather than simply the interoperability of file formats.

While some cloud platforms can be configured to provide highly granular object-level security, implementing interoperability  across more than one platform is complicated.  In addition, such approaches rely on centralized security management and infrastructure connectivity. Current approaches struggle to support intermittent connectivity scenarios and local-first communications. (Could that virtual cup of tea be poured across Metaverse platforms when users are sitting next to each other but do not have an Internet connection?)  

Our more general example, described in Section~\ref{sec:prototyping}, is collaborative editing. Today, it occurs entirely within siloed, online Web applications, which do not allow multiple editors to operate on the same hypermedia data in locally-connected scenarios.  Radoff describes {\bf interoperability} on different layers (connectivity, persistence, presentation, meaning, and behavior) \cite{Radoff_2022}
as a major challenge for future Metaverse systems, and discusses the importance of {\bf composability} \cite{Radoff_2022b},
alluding to composable computing and distributed computing as an example.  ARENA \cite{pereira2021arena} is an example of a {\bf modular system design} for multi-user and multi-application setups, enabling the development and hosting of collaborative XR experiences on WebXR-capable browsers with transparency, allowing data to migrate seamlessly  across computing resources.

This paper proposes to re-imagine the Metaverse (or, really, revisit the Web's original promise) as a data-oriented hypermedia system that enables \emph{secure, object-level interoperability for interactive applications} and service composability within and among applications, both without requiring intermediation by central server platforms for user authentication and message relaying. 
Our contributions can be summarized as follows.
First, we developed the {\bf Secure Web Object} (SWO) concept as a fundamental building block that enables securing web objects directly, independent of the connections they are conveyed in. 
Second, we conducted {\bf experiments by developing prototype applications using SWO}, which allowed us to obtain new insights into the feasibility and challenges of this approach. 
Third, based on our experience we formulated a {\bf research agenda for future work} in this direction.

The remainder of this paper is organized as follows. In Section \ref{sec:newpath}, we further discuss the opportunities in more detail, which are illustrated by a description of two prototype designs in Section \ref{sec:prototyping}. 
We summarize  the insights and challenges gained from prototype development in Section \ref{sec:insights} and articulate a suggested research agenda in Section \ref{sec:research}.
Finally, we discuss related work in Section \ref{sec:RelatedWork} and conclude this paper in Section \ref{sec:conclusion}.

%% file: 2-path.tex
\section{A New Path}
\label{sec:newpath}



\begin{quote}
{\em Information systems start small and grow. They also start isolated and then merge. A new system must allow existing systems to be linked together without requiring any central control or coordination.} -- Tim Berners-Lee\footnote{\url{https://www.w3.org/History/1989/proposal.html}}
\end{quote}

Today's Web functions mainly as a data-oriented application layer, where data is identified using URIs and managed through REST primitives \cite{fielding00}, exchanged through HTTP requests and responses. 
The underlying communication is connection-oriented, using TLS/TCP or QUIC, and the security model is consequently connection-based: browsers authenticate servers to set up secure connections. Trust anchors are root certificates that are installed in web browsers and operating systems. It is worth noting that users are not authenticated in that layer -- they do not have universally valid identities in this system. 
Instead, when required, users are authenticated by their accounts that have been set up on \emph{specific platforms} at the applicaiton layer.

When layering the Metaverse on top of this system (as most current Metaverse implementations do), objects in the Metaverse can be uniformly linked, albeit incurring considerable system complexity and run-time performance costs: URIs contain DNS names, which need additional resolution infrastructure to map names to server IP addresses. The actual transfer of web resources requires a secure connection setup, i.e., a unique transport session for each client-server pair. The digital representation of these objects cannot be shared outside this specific session, i.e., when an object is to be shared or re-used, a new secure connection must be established. Afanasyev et al. provide a more detailed discussion of these problems \cite{10.1145/3011883.3011890}.

This reflects a semantic gap between the data-oriented application layer and the connection-oriented transport layer, which becomes more evident for Metaverse applications that extend the Web into 3D interaction and immersion. The earlier example of pouring virtual tea from a pot in one Metaverse platform into a cup in another illustrates this.  
In such systems, 3D objects and live media data can be created continuously from a variety of sources and consumed by a group of participants, making the connection-oriented underlying Web platform no longer match the data-oriented nature of applications. It also leads to inefficient duplicated client-server transmissions during shared local experiences--imagine an augmented reality tea ceremony, because objects cannot be shared directly. 

In our work, we seek to enable Web applications to create and exchange Web objects, accessed by URI-like names, without relying on the communication channel for security. By decoupling data exchange from secure channels, we can (re-)enable an array of different interaction styles, from individual Web object access across different applications to scalable multi-destination distribution.  The secure web object  approach aims to bring the security and transport semantics \emph{as implemented} in Metaverse platforms closer to how users and developers think about and describe interaction within virtual worlds. 

Our approach to SWO is inspired by and built on the design concepts of Named Data Networking (NDN)~\cite{ndn2014}, a new way to network applications, platforms, users and devices, that produce and/or consume data and by earlier content-based security concepts for the web \cite{10.1145/3011883.3011890}. 
By \emph{Web objects}, we refer to named, secured, immutable data objects that have semantic meaning within an application domain, composed of lower-level application data units (ADUs)~\cite{clark1990architectural} that can be individually accessed by their names.  Examples include file-like objects such as 3D models, dynamic data such as the transforms associated with instances of those 3D models, as well as other forms of media, such as groups-of-pictures (GOPs) in streaming video.  
We propose that  a data-oriented security model, in which each ADU is cryptographically secured (e.g., signed and/or encrypted by keys that also have URI-like names), yields \textbf{secure Web objects} that are channel and platform independent.  SWO can support important emergent features at the Web layer, as cryptographic operations are performed on the objects themselves, rather than the channel they are carried over, including authentication, authorization, binding to real-world identities, and data encryption. Access to secured Web objects  becomes location-agnostic: data can be stored, replicated, and accessed just by name. 
Securing data directly removes the reliance on secured client-server connections, enabling Web applications to communicate securely in local, decentralized contexts.

Exchanging SWOs \emph{by names} requires name-based rendezvous.
This rendezvous function can be achieved in multiple ways, e.g. through a single rendezvous server, or establishing connectivity between all parties, or connecting all parties via NDN.
Our prototype applications, described in Section~\ref{sec:prototyping}, takes the third choice and transports SWO over \emph{NDN forwarders} which forward SWO requests according to their names towards the SWO producers. 
They can make use of any available packet transport service, such as HTTP, QUIC, TCP/TLS, UDP, WiFi, or Bluetooth, with or without transport security. 
Please see Section~\ref{sec:functions} for more details.


Running SWO-based applications over an NDN network brings additional advantages, including intrinsic multipath support, in-network caching and replication, and leveraging wireless broadcast communication for efficient data multicast.  
This leads to applications that are less susceptible to disruptions and changes in the connectivity.  They are likely to have lower latency because of their ability to use any (and all) available connectivity.
When implemented over NDN, SWO can be viewed as a set of naming, security, and payload conventions that provide a thin abstraction layer above NDN packet-level specifics for web applications. It extends work on generalized object formats in NDN such as \cite{thompson2019ndn,gusev2019data}.


This data-centric communication approach unifies single-destination and multi-destination delivery and can dramatically simplify both server scalability and distribution infrastructure. E.g., server offloading through caching becomes an intrinsic feature, because SWO can be easily replicated in managed ways or opportunistically cached by network elements and/or other endpoints.  Because their security properties do not emerge from the channel that carries them, SWO can be used directly in decentralized applications over potentially insecure transport as well as more ``traditional'' means.   This approach enables the expansion of the practical possibilities of the Web:
\begin{itemize}
    \item application and platform interoperability can emerge directly from common SWO data and naming formats, with cryptographic key exchange used to provide access to data; 
    \item secure interaction between locally connected nodes can be readily achieved without global connectivity;
    \item moving from browser interfaces that replicate server silos (e.g., the tab) to layered, possibly 3D-structured UIs more suitable to immersion concepts, such as the Metaverse; 
    \item enabling such cross-origin access to elements from multiple ``applications'', while tracking  identity and provenance as described in \cite{10.1145/3623565.3623761}; and
    \item on-demand data transformations in the network, such as rendering an object containing a 3D scene description into 2D tile objects, or data compression/de-compression.
\end{itemize}

%% file: 3-prototype.tex
\section{Experiments with SWOs}
\label{sec:prototyping}

%

Using SWO as the basic building block, we applied the data-oriented approach described above to develop two Microverse application prototypes running in browsers. 
In this section, we first describe a small set of functions that are necessary to enable the exchange of SWO by applications; the same set of functions is used in both prototype developments.
We then describe each of these two application prototypes.

\subsection{Functional Supports for SWO}
\label{sec:functions}
To allow applications use SWO to exchange semantically named, secured objects,  solutions are required for the following three tasks:
\begin{enumerate}
\item Securing web objects requires producers to possess cryptographic keys to encrypt and sign objects and consumers to have valid keys for signature verification and decryption. 

\item Exchanging SWOs by names requires name-based rendezvous.

\item Data-oriented communications let consumers fetch objects, which in turn requires consumers to be informed of object availability in order to fetch promptly.
\end{enumerate}
We make use of \emph{Named Data Networking} (NDN) \cite{ndn2014, ndn2018} primitives and libraries, in particular the NDN libraries in TypeScript~\cite{NDNts}, to carry out the above three tasks.

To build security into applications, entities go through a standard NDN bootstrapping process before they start running applications, to obtain a set of security parameters including the trust anchor, an entity's certificate, and security policies~\cite{2023cornerstone}.  
The first step in this bootstrapping process is new entity authentication: assuming that a user or device possesses an Internet-based name, such as an email address or DNS name, the bootstrapping process verifies the ownership of that identifier.
This step can also use different methods based on specific application scenarios.

The previous section listed several different options to enable the exchange of SWOs by names. Our prototype implementations interconnect all the SWO entities via NDN. Browsers are interconnected through simple schema-agnostic Websocket servers that perform NDN forwarding functions.
Object requests are forwarded to producers of the requested objects based on names, and the requested objects are returned to the requesters.
One can run a Websocket server locally, or run over an NDN-connected overlay. 
It is easy to set up one's own NDN overlay using the readily available containerized NDN routers\footnote{\url{https://github.com/named-data/testbed}}, or otherwise make use of the multi-continent NDN testbed, an overlay made of containers interconnected via TCP/UDP tunnels and providing global NDN connectivity. 
Supporting applications running over NDN does not require a global adoption of NDN. 

To inform all parties in a Microverse app about object availability, our prototype implementations make use of the NDN Sync protocol which promptly propagates new object production information to all parties~\cite{sync-sok}. 



\begin{figure}[htbp]
    \centering
    \includegraphics[width=0.48\textwidth]{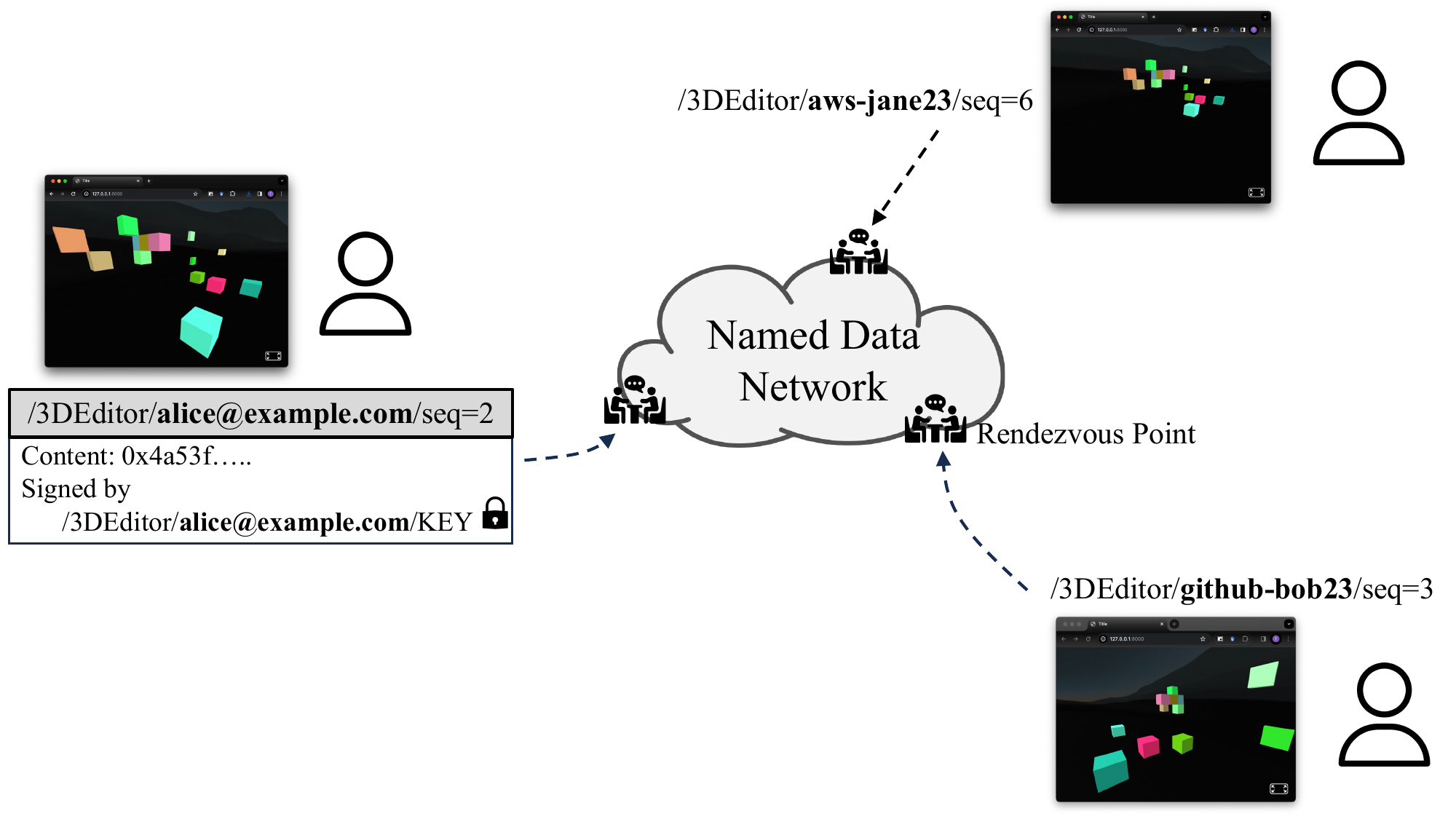}
    \caption{A collaborative 3D editing scenario among Alice, Bob and Jane, who have obtained their identifiers by adding the application prefix ``/3DEditor'' in front of their own identifiers.}
    \label{microverse-design}
\end{figure}

\subsection{Microverse Editor}
The first prototype that we developed, the \emph{Microverse Editor}, 
enables collaborative editing of 3D scenes expressed in SWO.
\footnote{see \url{https://named-data.net/microverse/} for a more detailed description of the project.} 
To jointly edit 3D scenes, the
\emph{Microverse Editor} lets users in the same web of trust directly exchange SWO with each other.  
The use of SWO enables objects to be stored anywhere, and for secure, collaborative editing without global connectivity. 
It also opens the possibility of different editor software operating on the same objects if entities have appropriate cryptographic keys.  

To join an existing web of trust, a user of the Microverse Editor uses their existing Internet identifier, such as an email address, cloud username, or DNS name, and produces a self-signed certificate.\footnote{While these identifiers come from global systems, once the certificates are created, they can be used even when there is no connectivity to the global Internet.}
Users can then mutually authenticate each other through direct or indirect trust relations, such as QR code exchanges or transitive trust relations.
As Figure~\ref{microverse-design} shows, scene updates are represented in SWO with URI-like names, and each user secures both assets and update objects by using his/her keys to sign the objects, with keys and certificates themselves as SWOs.
Therefore, SWO signing relations can be expressed using schematized name mapping onto application namespaces.
For example, one can easily define signing relations on SWO names that require all user changes made in this ``3DEditor'' be signed by corresponding user key SWOs.

Although our prototype \emph{Microverse Editor} has limited editing functions, its use of the SWO model enables several attractive features:
scene and object descriptions are expressed as individual SWOs and collections of SWOs.
A collection, such as a scene, can be seen as a manifest that refers to an individually named SWO, including other collections. 
Aggregation and re-use can easily be achieved by creating such collections. 
Because SWOs are uniquely named, copying an existing complex 3D scene is efficient. 

\subsection{NDN Workspace}
The second prototype we developed is a shared editor called \textit{NDN Workspace}~\cite{2024NDN-workspace},
\footnote{see \url{https://github.com/UCLA-IRL/ndn-workspace-solid}}, which shares the same set of supporting functions as the \emph{Microverse Editor}: assigning each user an identifier based on his/her unique existing Internet identifier, facilitating mutual user authentication, securing named data directly, and enabling the exchange of  named SWOs through a rendezvous place. 
When a local group of people jointly works on the same paper, \textit{NDN Workspace} keeps all the traffic local if an NDN forwarder is nearby.

One feature of NDN Workspace that the \emph{Microverse Editor} does not have (yet) is the capability to merge changes made by different users.
As multiple users may edit simultaneously, NDN Workspace must maintain a consistent ordering of all the changes seen by all users.
This goal is achieved by representing the files in a shared editor by a collection of Conflict-Free Replicated Data Types (CRDT)~\cite{shapiro2011conflict} data structures, which ensures consistent user views by achieving eventual consistency in shared data structures.
Workspace also allows users to work offline. Once offline users come online, they exchange SWOs with others to merge edits with the rest of the group.

\subsection{Preliminary Evaluation}
We performed a proof-of-concept experimental evaluation of both the prototypes.
In the Microverse Editor experiment, three users made changes to the shared 3D-scene through in-browser 3D-structure UIs, and exchanged SWOs over the NDN Testbed~\cite{ndn-testbed}.
The key locator in Figure~\ref{microverse-design} shows all three users the provenance of each SWO, and each user further checks SWO authenticity by verifying the signature with the corresponding key SWOs.

NDN Workspace has been in regular use since early 2024 for progress checking by a research group, where members of the group, which may be either local or distant, are NDN-connected through the NDN testbed.
The app has also been tested in scenarios with no Internet connectivity.
In this scenario, Alice and Bob jointly edited a file, exchanged SWOs, and then Alice went offline.
Bob then joined a local WiFI network without global Internet access and exchanged SWOs with Jane.
Although Alice was unreachable by Jane, Jane was still able to obtain Alice's SWO from Bob.
This is because SWO names are container-agnostic and secure, therefore anyone can fetch Alice's SWO by name from anywhere and verify the SWO's authenticity.






%% file: 4-insights.tex
\section{New Insights and Challenges}
\label{sec:insights}

Web and Metaverse applications are inherently data-oriented (on the application layer). The application layer data structures in Metaverse (e.g., 3D models and scene descriptions) are based on object hierarchies, in which connection-based systems cannot be fully used.
When we conceive `the Metaverse' not merely as an application running on the current network but as an evolution of the network itself, we can narrow rather than expand the gap between network architecture and application semantics.
Specifically, `the Metaverse' can be viewed as an information-centric system in which  applications engage in granular 3D content exchange, context-aware integration with the physical world, and other Metaverse-relevant services \cite{kutscher2023statement}. 
NDN generally facilitates direct data-oriented communication, enabling access to granular, individually secured objects, e.g., making up a video stream, directly by name as required by applications, independent of channel-based abstraction.
NDN implements request-response semantics at the network layer, akin to web semantics, but with packet-level granularity, operating without host addressing or name-to-address mappings such as those used in the Domain Name System (DNS).

The vision of a data-oriented Metaverse Web is based on data immutability as a fundamental, universal concept in facilitating the Metaverse (Web) as a distributed hypermedia information system. We argue that aligning hypermedia object models and access methods as well as the underlying network transport better with the inherent data-oriented application structure has important advantages with respect to ease of software development (both within and outside of the `app' paradigm), scalable sharing and multi-destination distribution, decentralized and asynchronous communication, and data re-use.

While Web servers will continue to play important roles in applications, we have shown that secure Web objects can support other communication models with additional advantages, such as decentralized editing of shared documents without requiring a central server. 
Our approach makes data objects first-class citizens of the Web again
and has security anchored to data instead of data containers and communication channels.
The use of semantic identifiers helps develop decentralized security solutions based on trust relations between users, instead of being determined by 3rd party (the current WebPKI security model).




%% file: 5-agenda.tex
\section{Research Agenda}
\label{sec:research}

Our prototype development demonstrated the potential of the SWO approach. Based on related ongoing research in named data networking~\cite{ndn2014}, we suggest the following research agenda towards creating a general SWO Web platform. 


\subsection{Naming}
Enabling distributed data-oriented applications requires {\bf namespace design strategies} for application domains, i.e., design strategies for developing name hierarchies that best express an application domain's semantics and  support its security needs. When considering interoperability between applications enabled by shared data access, the namespace design challenge becomes even more relevant: for example, it will be important to {\bf map common file and interchange formats intro named data} in ways that promote cross-application compatibility and increase the granularity of access and security. Hierarchical scene graphs in different 3D graphics and Metaverse contexts appear to lend themselves to fairly direct mapping to hierarchically named data, but present challenges in both design and engineering, including how to enable both efficient access to individual objects and multiparty sync of a dynamic scene. In general, more work is required to elaborate the {\bf analogies and technical mappings that are possible between Web semantics and named data}.

\subsection{Security}
Using SWO also requires easy-to-use data-centric security in which trust relationships can be described using names~\cite{nour2021access} and bootstrapped in a variety of ways. In the prototypes described in Section \ref{sec:prototyping}, we developed a pragmatic approach for {\bf security bootstrapping} in local networks or closed user groups, i.e., obtaining cloud-independent identifiers and security credentials. This is certainly an area that needs more attention, including initial deployment approaches for the development and deployment of SWO applications in today's Internet and Web environment, as described in \cite{2023cornerstone}. 
Another interesting area is the development of {\bf group security solutions} for data confidentiality, authenticity and access control in an SWO framework. 
In performance-critical applications, public-key-based security may not be the first choice, and technologies such as broadcast security can be leveraged. Additional requirements for supporting intra- and inter- application security (e.g., securely identifying creators of artifacts across application boundaries and securely controlling access to certain objects), for example through name-based access control, should be considered. This would provide fine-granular authorization on the SWO level and thus go beyond what current Web and Metaverse systems can do.


\subsection{Interactions}
With the foundation of the channel-agnostic exchange of named, secured objects, we can {\bf rethink application layer communication semantics and interaction styles}.
In today's Web, the client-server model has led to representational state transfer (REST) as a universal vehicle for any form of data exchange, leading to unnecessary complications for applications that do not require common state evolution between a client and a server. In a fundamentally data-oriented Web, simple publishing and access semantics can be realized consistently across a variety of situations. 

Knowing the data naming and security approaches is sufficient for publishing data or synchronizing data sets among inter-operating applications. For example, one might be able to publish a 3D scene by adding appropriately named, signed, and encrypted SWO into a namespace that others are watching for changes (or are notified about). APIs for publishing and access could also be agnostic to the number of peers because data-oriented communication has no concept of individual connections or peers. In NDN, there are implementations of so-called distributed dataset synchronization schemes that can be perceived as a multi-party transport layer for SWO communication. Client-server applications with robust state evolution can still be achieved through data-oriented REST approaches, such as those described in \cite{10.1145/3517212.3558089}.

With semantically meaningful identifiers,  interoperability can be facilitated and security properties and access control rules can be linked to application-defined names and name prefixes. Further work is required to {\bf explore different options for naming and data re-use}. For example, data re-use should be enabled across application boundaries, which suggests a level of indirection for naming. Such approaches would benefit from collection concepts for file-like collections such as FLIC \cite{I-D.draft-irtf-icnrg-flic} {\em and} dynamic objects.

\subsection{Application Frameworks}
Based on such abstractions, we can also explore how {\bf serialization to secure Web objects can be integrated into file systems, databases, and other storage mechanisms}, and how the interaction between server-side logic and in-browser based counterparts can benefit from such features. For example, reactive programming concepts (as used by libraries such as React\footnote{\url{https://react.dev/}} have led to the development of application platforms such as Next.js\footnote{\url{https://nextjs.org/}} and Remix\footnote{\url{https://remix.run/}} which enable end-to-end application development based on a reactive (partly functional programming) paradigm. Newer systems such as Electric Clojure\footnote{\url{https://github.com/hyperfiddle/electric}} go a step further and fully abstract  client/server state synchronization at the programming language layer, to achieve a strong composition across the frontend/backend boundary in dynamic Web apps. We believe that such systems can benefit from an SWO-based abstraction and native communication layers. This could be especially  useful for mobile application development platforms, such as Firebase\footnote{\url{https://firebase.google.com/}} which can benefit from the above-mentioned efficiency gains by mapping SWO directly to named data on the network.

Additionally, computation on data (`in-network computing') becomes easier to integrate, for example by a named-function design where functions create new (location-independent) secure Web objects following the same principles.  Decentralized communication and processing are readily available for any Web application.

\subsection{Communication}
Although we can tunnel SWO communication over existing QUIC or TCP and rely on relay servers, better performance and scalability properties can be achieved by direct peer-to-peer local SWO exchanges without server inter-mediation as in our prototypes.
We want to {\bf enable direct user-to-user SWO exchange on top of today's available communication underlays}, i.e., Internet protocols. We do this in our prototypes, running SWO implemented in NDN and communicating over both local connections and our global testbed. 

Further benefits can be gained by enabling the underlying NDN communication to occur over a range of transports, including lower-level protocols.  More work is needed on bootstrapping such communication, which is often hindered by the security sandboxing implemented in browsers owing to the limitations of connection-based security. For example, distributing SWO as network layer data objects would enable the use of wireless broadcast/multicast, which is currently not achievable for Web communication but could provide significant scalability benefits. Further work is required to enable this in real-world networks.

%% file: relatedwork.tex
\section{Related Work}
\label{sec:RelatedWork}


\subsection{Early Information-Centric Networking Systems}
\label{Information-Centric}

Early Information-Centric Networking Systems, such as NetInf \cite{10.1016/j.comcom.2013.01.009}, proposed an \emph{information layer} that could be used to access objects by names, independent of location.
These names would then be layered on top of a \emph{``bit-level'' transport layer} to be used for the actual data transfer from a specific location via a specific protocol. 
Such systems were inspired by P2P networking, and required \emph{name resolution} service to map an object name to a \emph{transport identifier}. 
Simple systems often used DHTs such as Kademlia \cite{maymounkov2002kademlia}, while more elaborate systems such as NetInf used hierarchical multi-layer DHTs, namely MDHT \cite{10.1145/2018584.2018587}.

From an architectural perspective, in such name-resolution-based designs,
the namespace for named data objects differ from that used in the communication network, which is often locator-based. 
Although it is relatively easy to run over the existing Internet, such systems would have strong dependencies on both 
(1) the name resolution infrastructure for bootstrapping any communication, which would rule out decentralized, local-first communication scenarios; and
(2) the underlying IP network infrastructure, as they themselves do not have a routing/forwarding layer. As such, they would also be unable to leverage data-centric forwarding features, such as in-network replication, caching, and loss recovery.

\subsection{Social Linked Data (Solid)}
\label{Solid}

















Solid \cite{sambra2016solid} is a linked data-based decentralized platform for social Web applications. It aims to provide users with decentralized data storage, granting users full control over their own data, including access control and storage location.
In Solid, each user stores their data within an online storage space known as a personal online datastore (pod). Pods are web-accessible storage services that users can either host on personal servers or entrust to public pod providers, similar to current cloud storage providers (e.g., Dropbox) \cite{mansour2016demonstration}. 
Pod providers vary in their degree of privacy and reliability (e.g., availability or latency guarantees), or offered legal protection (e.g., the legal frame of the country hosting the pod). 
Users may have more than one pod and select among different pod providers based on their specific needs. 

Solid uses linked data \cite{berners2006linked} to achieve data management by assigning each piece of data a unique Uniform Resource Identifier (URI), which points to its location within a specific pod.
The URI includes the domain name of the pod and path to the data object. 
For example, \url{https://username.solidpod.provider.com/data/mydata.ttl} is a URI where:
\url{https://username.solidpod.provider.com/} is the base URI for the user's pod, and \url{/data/mydata.ttl} is the path to the specific data object within the Pod.
Data operations are then conducted using RESTful HTTP methods.
Users add new data items to a container either by sending them to the container's URL with an HTTP POST or by placing them within its URL space using an HTTP PUT. Updates are made using HTTP PUT or HTTP PATCH, deletions with HTTP DELETE, and retrieved using HTTP GET and following links.



Unlike the current Web, Solid provides its own user authentication and authorization.
Initially, users create an identity by registering with an Identity Provider. This identity is typically represented by a WebID, which is a URL that points to a \textit{profile document} describing the user and their public data (e.g., \url{http://somepersonalsite.com/#webid}). During the authentication process, users validate their identity through the Identity Provider using their WebID via a decentralized authentication protocol called {\em WebID-TLS}.
WebID-TLS allows users to authenticate themselves on any site by choosing one of the client certificates offered by their browser. Unlike classic client certificate authentication that depends on Public Key Infrastructure (PKI), these certificates do not need to be issued by a trusted Certificate Authority. 
Instead, 
each client certificate includes a field called Subject Alternative Name, which contains the user's WebID, thereby linking the certificate to the user’s WebID. During the authentication process, verifiers only need to match the certificate’s public key with those listed in the \textit{profile document} associated with the WebID. From the user’s perspective, the complete  WebID-TLS authentication process is a one-click operation for choosing the WebID certificate.
Once the identity is verified, the Identity Provider issues a token (often a JSON Web Token (JWT)) to certify the user's identity for the duration of a session.
Upon attempting to access the resources stored in a pod, users send a request along with a token received from the Identity Provider. The server hosting the pod then verifies the token to check its validity and determines whether the authenticated user has the necessary permissions to access the requested resources.






In summary, Solid provides two main new features over the existing web: decentralized storage and service hosting, and a user authentication and authorization scheme based on WebID and Web Access Control relying on bespoke identity profile servers.
However, communication within Solid uses the existing Web protocols
and is still connection-oriented, and URIs are Pod-specific, i.e., tied to specific servers. Data-oriented applications would still need to bridge the gap with server-centric communication, thus functions such as direct user-to-user communication and efficient object sharing are still not directly supported, nor are low-latency, local-first communications.

\subsection{Web 3.0}

Web 3.0, or the decentralized web~\cite{Browne_2022}, aims to transform the existing Internet infrastructure by adopting a decentralized, distributed architecture that is both trustless and permissionless. `Trustless' refers to the design of the network, which allows participants to interact either publicly or privately without relying on a trusted intermediary.
`Permissionless' access means that all parties, whether users or providers, can engage with the network without requiring approval from a centralized authority.


One instantiation of Web 3.0 is the InterPlanetary File System (IPFS).
IPFS is a blockchain-enabled, content-addressable peer-to-peer network that provides distributed data storage and delivery \cite{enwiki:1219603182}.
Compared to Solid (Section \ref{Solid}), which focuses on controlling data usage and access, IPFS has a broader focus on providing a decentralized storage system, e.g., by serving as a back-end \cite{parrillo2021solid}.
At its core, IPFS incorporates four main concepts \cite{trautwein2022design}: I) content-based addressing: unlike HTTP, IPFS detaches object names from their host location , allowing data to be accessed from any node; II) decentralized object indexing: IPFS employs a decentralized peer-to-peer overlay network to index all available locations from which objects can be retrieved, making the system less vulnerable to failures. Instead of devising an entirely new system, IPFS uses the Kademlia DHT (see Section \ref{Information-Centric}) for content indexing \cite{maymounkov2002kademlia}; III) immutability and self-certification: by using cryptographic hashing, IPFS ensures the immutability of objects and self-certifies their authenticity, obviating the need for traditional certificate-based authentication and enhancing verifiability; and IV) open participation: the network allows anyone to deploy an IPFS node and participate without needing special permissions or privileges.


Similar to BitTorrent \cite{BitTorrent} or information-centric concepts (discussed in Section \ref{Information-Centric}), IPFS employs a content-based addressing scheme using unique hash-based Content Identifiers (CIDs) \cite{IPFS}. CIDs are base primitives that decouple content names for their storage locations. By contrast, location-based systems (e.g., HTTP) bind content addresses (i.e., URLs) to their primary hosts. This design facilitates decentralization of content storage, delivery, and address management. Moreover, by decoupling the content address from its storage location, CIDs help eliminate vendor lock-in and reduce reliance on central authorities for address allocation.
In fact, IPFS appears to be a limited variant of earlier information-centric systems, such as NetInf. While these proposals primarily implement content-based addressing at the network layer, IPFS relies on application-layer routing. 
IPFS operates at a coarse-grained level, which can be less efficient for handling small or frequently updated files that are common in modern web applications.
Currently, the IPFS network is used more as a file transfer system than decentralized storage owing to the short lifetime of data items. Storage persistence relies predominantly on a few centralized, cloud-based providers. 
Filecoin \cite{trautwein2022design}, a decentralized and incentivized storage network built on top of IPFS, is intended to enhance storage persistence as a content storage incentive and monetization mechanism \cite{balduf2023cloud}.
However, it is unclear whether decentralized storage nodes can compete with dedicated cloud providers in terms of  service quality and cost.
Additionally, data retrieval in IPFS is generally slower than in direct HTTP access. Data may need to be fetched from nodes that are geographically distant or have slow connectivity.

Moreover, in reality, IPFS is not as centralized as claimed. Recent studies \cite{balduf2023cloud} have indicated a high degree of traffic centralization in the IPFS network, with a major portion of nodes hosted in the cloud. Notably, the top 5\% of nodes handle up to 95\% of network traffic, with Amazon-AWS (one of the largest cloud providers) generating 96\% of all content resolution requests. 
They also highlighted the heavy reliance on cloud infrastructure for content storage. Approximately 95\% of the content is hosted by at least one cloud-based node, with cloud-based storage platforms using IPFS (e.g., web3-storage and nft3-storage) holding a major share of the persistent content in the network. Although this setup enhances convenience and accessibility for end users, it significantly contributes to the centralization of the network.


In summary, IPFS appears as a decentralized P2P storage network, providing similar functionality to BitTorrent and early Information-Centric Networking systems, but with less scalability and performance. It appears to be a use case for monetization through Filecoin and less a viable platform for the future Web, let alone the Metaverse.

%% file: conclusion.tex
\section{Conclusion}
\label{sec:conclusion}
Variations in the concept of named, secured data objects already exist within individual applications, particularly for those that cannot rely solely on channel security.
We propose that adopting consistent naming, security, and access mechanisms to create an interoperable, channel security-independent approach for secure exchanges of data opens up important new possibilities. When cryptographic operations are performed on the objects themselves, access to web objects and dynamic computation results becomes location-agnostic, enabling the flexible use of path diversity, different modes of transmission (point-to-point and multi-destination), as well as opportunistic and managed in-network storage. 
Uniform Resource Identifiers (URIs) would name SWOs independent of their location and access modalities (as intended by the original URI concept), and object authenticity is based on the names directly. We have shown how these concepts can enable the design of new types of Metaverse and Web applications, by supporting both traditional RPC-like requests-response interactions as well as dataset synchronization with eventual consistency and CRDT-based state evolution. We can enable both traditional ``client-server'' interactions as well as decentralized "local-first" communication scenarios.

Metaverse is a particularly demanding and feature-rich Web application. We believe that the SWO concept would not only benefit Metaverse but also present a new way to evolve the Web at large. 
SWO-based systems require name-based rendezvous services. In traditional systems, this is typically performed at a central server that can recognize SWO names. An NDN network performs this rendezvous function inside the network by having requests carrying names meets named objects, creating a uniform framework to support both large-scale content distribution and decentralized, ``local-first'' communication. This new framework would facilitate application development and reduce operational cost.

The SWO approach should not be confused with other approaches that are sometimes referred to as {\em web3},
such as IPFS. While claiming to provide a platform for the next Web, these systems merely provide a (conceptually) decentralized storage platform for accessing file-like objects, which is neither a useful service model for future Web, nor enabling the data-oriented benefits as described here. While the SWO approach can in principle leverage existing transports as underlays, it is important to note that a direct mapping to named-data oriented communication would provide significant benefits, such as avoiding combinatorial interoperability problems and enabling efficient and local communications.

The SWO concept is complementary to Linked Data approaches such as the one proposed by the Solid project. Solid aims to link data hosted at decentralized data stores which are essentially web servers that are accessed via existing HTTP and transport protocols over IP.
As such, Solid inherits the current Web security model and its connection-oriented transport schemes. We believe that the Solid concepts can be implemented on top of an SWO platform, mapping objects in Solid pods to SWOs. We will analyze the security implications in future work.

Hence, in this paper, we present an early concept for secure Web objects that leverages research in data-centric networking to help the Web evole into a more effective environment for local-first, decentralized, and interoperable applications that resurface the early promise and excitement of the Web.  


%% file: 8-acks.tex
\section{Acknowledgments}
\label{sec:acks}

This work was supported in part by the US National Science Foundation under award 2019085 and 2126148, and the China Guangzhou Municipal Key Laboratory on Future Networked Systems (024A03J0623).  The views and opinions expressed in this study are those of the authors and do not necessarily reflect the official policy or position of our sponsors.